\documentclass{blois}

\newcommand{\C}[1]{{\cal C}_{#1}}

\newcommand{\av}[1]{\langle #1 \rangle}

\def\be{\begin{equation}}
\def\ee{\end{equation}}
\def\bea{\begin{eqnarray}}
\def\eea{\end{eqnarray}}

\newcommand{\Photo}{
}

\begin{document}
\vspace*{4cm}
\title{STATUS of NEW PHYSICS IN CP VIOLATION AND RARE B DECAYS}

\author{ JOAQUIM MATIAS}

\address{Grup Fisica Teorica, Department de Fisica, UAB-IFAE, 
E-08193 Bellaterra. \\ \medskip {\bf  Talk presented at 29th Rencontres de Blois, Loire Valley, France, June 2017}}

\maketitle\abstracts{
We briefly report on some particular aspects  of  new physics searches concerning, on the one side, the CKM paradigm and, on the other, the global fit to rare $b\to s \ell\ell$ observables including lepton flavour universality ones. We put special  emphasis on the state-of-the-art of hadronic uncertainties of $b\to s\mu\mu$ observables  and LFUV in SM and in presence of New Physics. Finally, we discuss the latest experimental and theoretical developments concerning long-distance charm contributions.
}

\section{Motivation}

The central question of Quantum Field Theory based particle physics is which is the lagrangian that describes fundamental interactions and associated to this question is:  which are the degrees of freedom, symmetries and scales. The Standard Model (SM) is up to now the best answer but it leaves many open  questions like the origin of dark matter, dark energy, the baryon asymmetry of the universe, etc. All those questions call for a more fundamental theory beyond the SM.


Processes governed by Flavour-Changing Neutral Currents are used systematically as a  tool to test and try to understand the flavour structure of the underlying fundamental theory. Also the CKM paradigm in the SM has been under intense scrutiny.
In this  proceeding we start in Sec.~2  very briefly reporting the few tensions in the CKM sector but the main focus is devoted in Sec.~3 to the systematic and coherent deviations observed in $b\to s \mu\mu$~\footnote{Unfortunately the interesting LFUV observables in $b \to c \ell \nu$ governed transitions covered in the talk are not discussed here due to editorial constraints. In this case we address the reader to the talk or to other proceedings.} and universal lepton flavour observables and the discussion of their  associated uncertainties.

\section{Asesssing the CKM paradigm in the SM}

In SM weak charged transitions mix quarks of different generations. This is encoded in the unitary CKM matrix \cite{ca,ko} that involving 3 generations allows for 1 phase as the only source of CP-violation in the SM.  
One of the off-diagonal equations coming from the constraint $V_{\rm CKM}^\dagger V_{\rm CKM}= 1$
defines a non-squashed triangle, that is usually referred as the Unitary Triangle (UT). The angles of this UT  are denoted by $\alpha$, $\beta$ and $\gamma$. 
The elements of the CKM are extracted from tree and loop processes (see \cite{seb} for a detailed discussion), with the corresponding uncertainty associated to the uncertainties of the decay used. For instance for $|V_{cb}|$ ($|V_{ub}|$) the main uncertainty comes from Form Factors (FF) of $B \to D^{(*)}\ell\nu$ ( $B\to \pi \ell\nu$) respectively. 
Bag parameters $B_K$ ($B_{Bd}$) are also an important source of uncertainty to determine the UT angles $\beta$ ($\alpha$) respectively. 

The systematic way to search for New Physics (NP) in the CP-violation sector is to overdetermine sides and angles of the UT. The result of this analysis using a frequentist approach (CKMfitter \cite{seb}) is that no significant deviations are observed when comparing determinations of the UT using CP conserving only observables, CP-violating ones,  tree or loop governed processes. 

Still there are two persistent tensions  between exclusive and inclusive determinations of $V_{ub}$ and $V_{cb}$. The tension in $|V_{cb}|$ between lattice determinations of exclusive decays in $B \to D^* \ell \bar{\nu}$ \cite{aoki} and inclusive ones \cite{gamb,amhis} is at the 3$\sigma$ level. However if sum rules \cite{gamb2} are used for the exclusive the tension vanishes. Finally indirect fits using CKM, CP violation and flavour data (except direct decays) gives a determination for $|V_{cb}|$ \cite{ckm2} closer  to the inclusive. The situation for $|V_{ub}|$ is different, here the inclusive determination \cite{amhis} is more challenging than in the previous case and the indirect fit \cite{ckm2} is more consistent with the exclusive determination \cite{aoki}. The tension is also at the level of 2-3$\sigma$ but if ${\cal B}(B^+\to \ell^+\nu_\ell)$ is used (averaging Babar and Belle) \cite{pedra} to determine $|V_{ub}|$ the result is consistent with both given the large error bar, not yet competitive (see also \cite{Kang}).

A natural question is if these tensions can be solved in the context of NP. And the answer according to \cite{crivellin} using an effective field theory  approach is that it seems not to be the case. In \cite{crivellin} it is shown that four-fermion operators generated at tree level like: ${\cal O}_R^S=\bar{\ell} P_L \nu \bar{q} P_R b, \,  {\cal O}_L^S=\bar{\ell} P_L \nu \bar{q} P_L b$ and  ${\cal O}_L^T=\bar{\ell} \sigma_{\mu\nu} P_L \nu \bar{q} \sigma^{\mu\nu} P_L b$ with $q=u,c$ always give a larger contribution to the inclusive than to the exclusive which makes impossible to reduce the distance between both determinations. A second possibility modifiying $W-qb$ couplings (generated via loop) 
has better chances to find common solutions but at the price of inducing tensions in $B \to \tau \nu$ and also generate a too large $Z-b\bar{b}$ coupling.




\section{Probing New Physics via Rare B decays}

The framework of an effective Hamiltonian to describe the $b\to s \ell\ell$ transition  allow us to separate short and long distances, with short distances encoded in the Wilson coefficients of the relevant operators and long distance in the matrix elements of these operators.
The relevant operators in the present discussion are:
\begin{itemize}
\item ${ {\cal O}_7} = \frac{e}{16 \pi^2}m_b\, 
        \bar s\sigma^{\mu\nu}(1+\gamma_5)F_{\mu\nu}\,b$ \quad {\small [real or soft photon]}
\item ${ {\cal O}_9}=\frac{e^2}{16 \pi^2}\bar{s}\gamma_\mu(1-\gamma_5)b\  \bar\ell\gamma^\mu\ell$
 \   {\small [$b\to s\mu\mu$ via $Z$/hard $\gamma$\ldots]}
\item ${ {\cal O}_{10}}=\frac{e^2}{16 \pi^2}\bar{s}\gamma_\mu(1-\gamma_5)b\  \bar\ell\gamma^\mu\gamma_5\ell$
\quad  {\small [$b\to s\mu\mu$ via $Z$]}
\end{itemize}
NP either induces a change in the Wilson coefficients by adding a new contribution or generate new operators (chirally flipped, scalar or pseudoscalar or tensor operators). 
We extract the information on Wilson coefficients  by means of a global analysis \cite{global,pattern} based on a frequentist approach including a set of 175 observables  in total (each bin corresponding to an observable) coming from LHCb, Belle, ATLAS and CMS experiment:

\begin{itemize}
\item $B\to K^*\mu\mu$ ($P_{1,2},P'_{4,5,6,8}$ \cite{piobserv0,piobserv1,p5p,piobserv2}, $F_L$ in 5 large-recoil bins + 1 low-recoil bin)+available electronic observables.
Also include April's update of ${\rm Br}(B\to K^*\mu\mu)$ \cite{brlhcb} showing now a deficit in muonic channel and
April's new result from LHCb on $R_{K^*}$ \cite{rkstarlhcb}.

\item $B_s \to \phi \mu\mu$~\cite{bsphimumu} ($P_1,P'_{4,6},F_L$ in 3 large-recoil bins + 1 low-recoil bin)
\item $B^+\to K^+\mu\mu$, $B^0\to K^0\ell\ell$ (BR) ($\ell=e,\mu$) ($R_K$ \cite{rk} is implicit)
\item $B\to X_s\gamma$, $B\to X_s\mu\mu$, $B_s\to \mu\mu$ (BR).
\item Radiative decays:
$B^0\to K^{*0}\gamma$ ($A_I$ and $S_{K^*\gamma}$), $B^+ \to K^{*+}\gamma$, $B_s \to \phi \gamma$
\end{itemize}
But also the important  Belle results \cite{wehle} on the observables $Q_{4,5}$ \cite{asessing} constructed from the isospin-averaged but lepton-flavour dependent observables are included:
\begin{equation}
P_{i}^{\prime\,\ell} = \sigma_+\, P_{i}^{\prime\,\ell}(B^+)
+ (1-\sigma_+)\, P_{i}^{\prime\,\ell}(\bar B^0)\ \nonumber
\end{equation}
We take into account the effect of the isospin average by means of the parameter $\sigma_+$:
This parameter in absence of more information is treated as a nuisance parameter $\sigma_+=0.5 \pm 0.5$. Finally ATLAS \cite{atlas} result on the full set of $P_i$ observables and also the two partial CMS analysis with 7 and 8 GeV (only a subset is measured) are included. In the case of the CMS \cite{cms} analysis it would be very interesting to confirm the stability of those results by extracting the three observables $F_L$, $P_1$ and $P_5^\prime$ altogether, as all other experiments did and not separately as  CMS does. This is crucial to make sure that no correlations affecting the extraction are being lost.

We use a wide set of tools depending on the observable used and the region in dilepton invariant mass for the semileptonic decay: OPE for the inclusive, QCD factorization including $\alpha_s$ \cite{alphas1,alphas2} and power corrections factorizable~\cite{power} and non-factorizable~\cite{hadronic,KMPW} for exclusive semileptonic at large-recoil and heavy quark effective theory, lattice QCD and quark-hadron duality at low-recoil.

\subsubsection{Several tensions  and two different types of anomalies observed}

In the large set of observables mentioned above clearly two type of anomalies are being observed systematically (some of them in different experiments): \begin{itemize}
\item Type-I: Anomalies associated to $b \to s \mu^+\mu^-$ transitions.
\begin{itemize}
\item $P_5^\prime$ observable \cite{p5p}. This is an observable whose dependence on soft FF cancel at LO, $P_5^{\prime}=P_5^{\prime \infty} (1+ {\cal O}(\alpha_s \xi_\perp)) + p.c.=$. This property unique to the optimized observables (other observables like $F_L, S_i= {\cal O}(\xi_\perp/\xi_\|)$ \cite{si} or ${\cal B}_{B \to K^*\mu\mu}={\cal O}(\xi_\perp^2, \xi_\|^2)$ are unprotected) reduce their sensitivity to the FF's choice. This can be easily seen in Fig.1 where  two completely different theoretical approaches using also different sets of FF (KMPW~\cite{KMPW} or BSZ~\cite{BSZ}) find results for this observable that are in excellent agreement. This is one of the most tested anomaly, it was observed a 3.7$\sigma$ deviation by LHCb \cite{ffi} in 2013 in a long bin, confirmed by LHCb in 2015 with deviations of $\sim 3\sigma$ in two adjacent bins and also by Belle  at 2.6$\sigma$ level  \cite{belle2} in the same region.

\item ${\cal B}_{B_s \to \phi\mu\mu}$ shows deviation at $\sim 2.2 \sigma$ level \cite{bsphimumu} in both large and low-recoil bins. 
\item ${\cal B}_{B^+ \to K^{+*}\mu^+\mu^-}$ also exhibits a 2.5$\sigma$ tension in the long low-recoil bin [15,19] GeV$^2$.

\end{itemize}

\item Type II: Anomalies in lepton flavour universality  observables (LFU), namely in ratios $R_{\rm P,V}={\rm BR}(B \to {\rm[P,V]} \mu^+\mu^-)/{\rm BR}(B \to {\rm[P,V]} e^+ e^-)$~\cite{rk,rkstarlhcb} and $Q_i$~\cite{wehle}. These observables introduce a new important element, namely, the first hints that Nature seems to violate the predicted universality of the SM of the leptonic flavour in these ratios.

\end{itemize}
But two crucial observations  in \cite{global,pattern} link both anomalies. First, a separated analysis of electronic observables gives a result compatible with no NP in the $C_i^{e}$,  still with very large error bars  but pointing in the direction that NP affects mainly $C_i^\mu$. And second and more important, in this case we found that the same NP explanation is consistent with both type-I and type-II anomalies. In Sec. 3.3 we will provide an explicit example of it.


\begin{figure}
\begin{minipage}{7cm} \vspace*{-0.3cm}
\includegraphics[width=7.5cm,height=4.5cm]{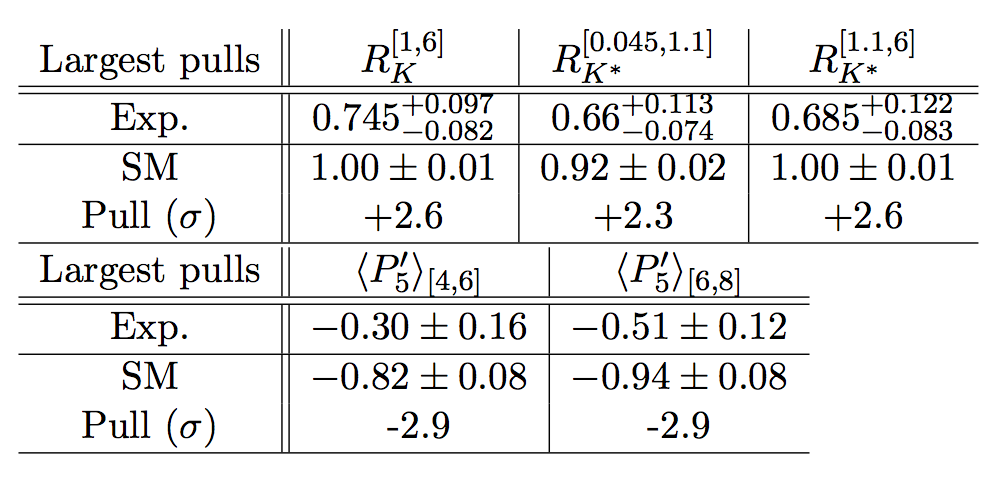}
\end{minipage}
\,\,\, \begin{minipage}{7cm}
\includegraphics[width=8cm,height=5.0cm]{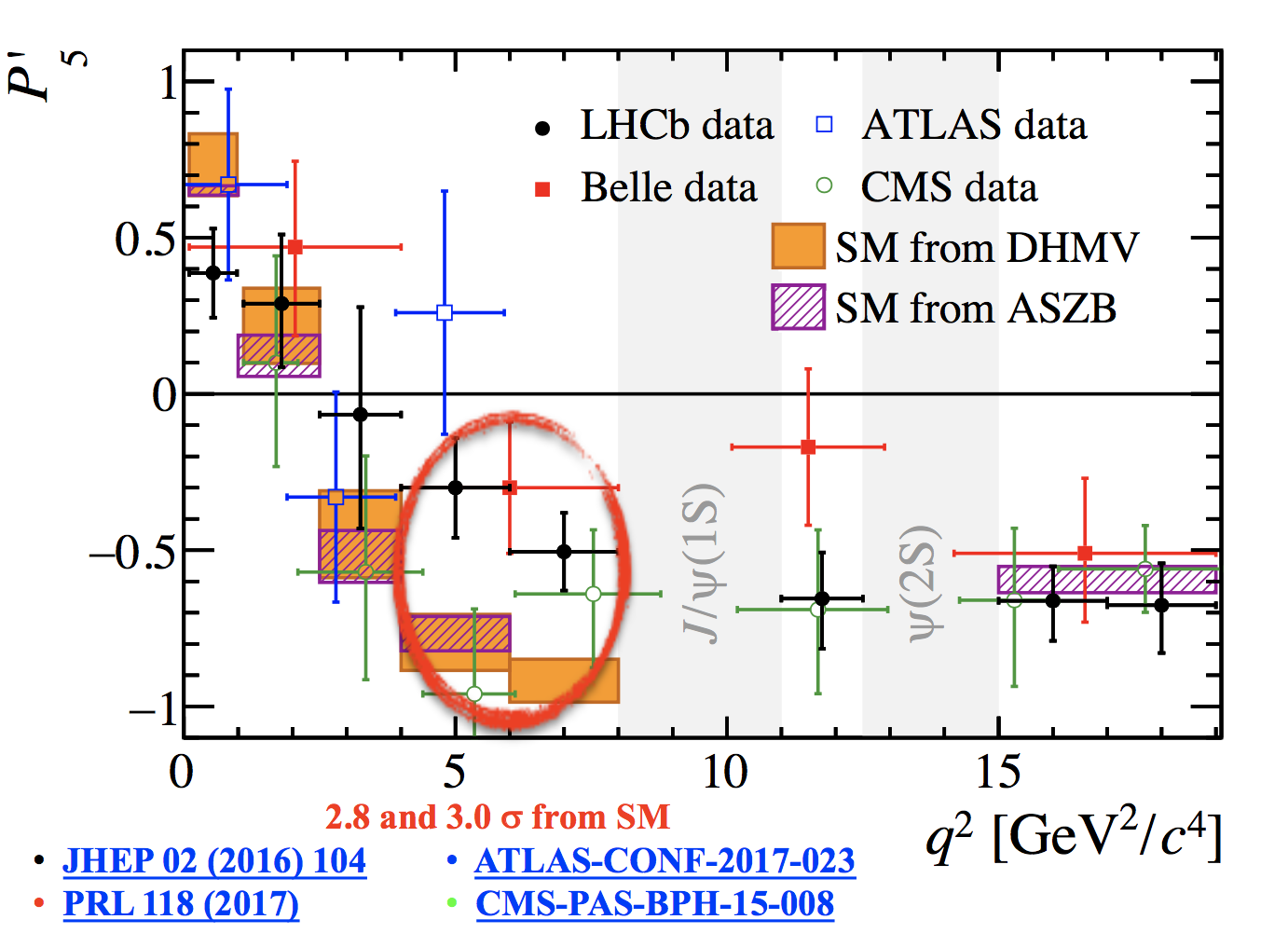}
\end{minipage}
\caption{Some of the main anomalies currently observed in $b\to s\ell\ell$ transitions and LFUV observables}\end{figure}

\subsection{Global fit result}
In \cite{pattern}  two types of analysis were presented, a complete analysis including the 175 observables mentioned above and a second one with only the subset of LFUV observables ($R_{K,K^*}, Q_{4,5}$). In both cases the set of 'radiative' and leptonic decays  are always included. The SM point yields a $\chi^2$ corresponding to a p-value of 14.6\%  for the first and 4.4\% for the second. The SM pull exceeds 5$\sigma$ (see \cite{pattern} for details) in the three main hypothesis of the complete analysis ($C_{9\,\mu}^{\rm NP}$ \cite{understanding}, $C_{9\,\mu}^{\rm NP}=-C_{10\,\mu}^{\rm NP}$ and $C_{9\,\mu}^{\rm NP}=-C_{9\prime,\mu}^{\rm NP}$), even if the last one is unable to explain $R_K$. But the most interesting outcome is that the 6-D fit (see Tab.1) has shifted, for the first time, to the 5$\sigma$ level from the previous 3.6$\sigma$. The fit to only LFUV observables exhibits a 4$\sigma$ significance in several of the NP hypothesis in front of the SM solution.
Finally the implications for models for each scenario are discussed in \cite{pattern}, but also in a very long (and incomplete) list of works (see for instance \cite{models2,models3,models4,models5,models6} and references therein).

\subsection{Hadronic uncertainties in $b \to s \ell\ell$ observables}

Let's focus for a moment on the optimized observables for $B \to K^*\mu^+\mu^-$. At large-recoil, where the most relevant tensions are observed, we work in the framework of improved QCDF using KMPW FF \cite{KMPW} as inputs and we include four types of corrections: known $\alpha_s$ factorizable and non-factorizable corrections within QCDF \cite{alphas1,alphas2} and two types of power corrections (p.c.), factorizable ones entering the decomposition of a FF and non-factorizable ones entering the amplitude coming from nonfactorizable soft-gluon emission from c-quarks \cite{KMPW}. It is important to emphasize that all those corrections are relevant both in the SM and in presence of NP. 
 
 Factorizable corrections have been discussed  at length in a series of works \cite{power,hadronic}  and the origin of the problem in estimating the errors in \cite{jaeger2} has been clearly identified. We will not repeat all details in this proceeding again but refer the reader to \cite{hadronic} where the state-of-the-art of hadronic uncertainties are discussed in full detail. In brief, factorizable power corrections ($\Delta F^{\rm p.c.}(q^2)$) appear in the decomposition of the full FF as the terms that break at order $\Lambda/m_B$ the symmetry relations among FF in an expansion of the dilepton invariant mass:
 $$F^{\rm full}(q^2)=F^{\rm soft}(\xi_\perp,\xi_\|)+ \Delta F^{\alpha_s}(q^2)+ \Delta F^{\rm p.c.}(q^2)$$
 The term $\Delta F^{\alpha_s}(q^2)$ instead  represents the breaking of the symmetry relations by known ${\cal O}(\alpha_s)$ corrections. A correct evaluation of the $ \Delta F^{\rm p.c.}(q^2)$ contribution requires  an accurate analysis, for instance, if the errors associated to the different power corrections are taken as uncorrelated then is compulsory to chose a scheme (a procedure to define the soft FF in terms of full FFs)  that minimizes the sensitivity to these corrections. Otherwise there is the risk of 
 inflating artificially the error size estimate like in \cite{jaeger2}. This is illustrated in Tab.~2 where the case with all correlations included (last row) is compared to the case with uncorrelated errors (first row) in two schemes (scheme 1 from \cite{global,hadronic} or scheme 2 from \cite{jaeger2}).\footnote{ Moreover, the authors of \cite{jaeger2} follow a rather unconventional procedure to assign an error to the soft FF. In \cite{jaeger2} the spread of central values of a small subset of FF (KMPW, BZ, DSE) determinations is used to fix the error of the soft FF but {\it not including any error associated to each determination}. As a consequence they get for $\xi_\perp=0.31 \pm 0.04$ (see Eq.12 in \cite{jaeger2}). This error in \cite{jaeger2} is factor of 5 smaller than the one we find  (and use in our predictions) $\xi_\perp=0.31^{+0.20}_{-0.10}$ taking as an input only one of those determinations (a very conservative one like KMPW) but, of course, including in our case  the error estimate associated to the corresponding FF. 
  The combination of both problems  (overvaluation of p.c. error and undervaluation of soft FFs one)  has the consequence that optimized observable errors are oversized while non-optimal observables like $F_L$ gets undervaluated its error in \cite{jaeger2}. This problem also leaks in LFUV observables estimates in \cite{jaeger3} in presence of NP. } This was proven numerically in \cite{power} and analitically in \cite{hadronic}. It was also found in \cite{hadronic}  that  the numerically leading term in a power correction expansion in $P_5^\prime$ (see eq. 21 in \cite{hadronic}) is missing in \cite{jaeger2}.

\begin{table*}[t]
\begin{center}
\begin{tabular}{c||c|c|c}
 & $\C7^{\rm NP}$ & $\C{9\mu}^{\rm NP}$ & $\C{10\mu}^{\rm NP}$   \\
\hline\hline
Best fit & +0.03 & -1.12 & +0.31  \\ \hline
1 $\sigma$ & $[-0.01,+0.05]$ & $[-1.34,-0.88]$ & $[+0.10,+0.57]$ \end{tabular} \\
\begin{tabular}{c||c|c|c}
 & $\C{7^\prime}$ & $\C{9^\prime \mu}$ & $\C{10^\prime \mu}$  \\
\hline\hline
Best fit &  +0.03 & +0.38 & +0.02 \\ \hline
1 $\sigma$  & $[+0.00,+0.06]$ & $[-0.17,+1.04]$ &$[-0.28,+0.36]$ 
\end{tabular}
\end{center}
\caption{1 $\sigma$ confidence intervals for the NP contributions to Wilson coefficients in
the six-dimensional hypothesis using a very large sampling. The SM pull is 5$\sigma$.}
\label{tab:Fit6D}
\end{table*}

 \begin{table}
\small
\centering
\renewcommand{\arraystretch}{1.5}
\begin{tabular}{@{}c|c|c@{}}
$\av{P_5^\prime}_{[4.0,6.0]}$ & scheme 1 & scheme 2  \\
\hline
a & $-0.72\pm 0.05$ & $-0.72\pm 0.15$ \\
\hline
full BSZ& \multicolumn{2}{c}{$-0.72\pm 0.03$} 
\end{tabular}
\caption{SM prediction for $P_5^\prime$ in the  bin $[4,6]$\,GeV$^2$ together with the error from soft FFs and factorisable p.c. (all other sources of errors have been switched off). Results  shown for: a) error of p.c. estimated $\sim F^{\rm LCSR} \times {\cal O}(\Lambda/m_b)$ + correlations by large-recoil symmetries and full BSZ (all correlations including LCSR ones in BSZ). 
Results displayed  in optimal scheme 1 used in our predictions (see [11,15,26])  but also in scheme 2 used in [40]. The example illustrates  that scheme 2 error is inflated  by a factor of 5 compared to  correlated case error (0.03) and that large-recoil symmetries are dominant
(see  sec 3.2 of [28]  for a detailed explanation  but in short while explicit computation in BSZ gives a 5\% size for p.c. scheme 2 corresponds effectively to an arbitrary $>20\%$). }
\label{tab:P5p}
\end{table}

Instead a real issue that was suggested was the possibility that long distance charm contributions could be the responsible behind some of the observed anomalies. The reason is that $\bar{c}c$ contribution in decays like $B \to K^* \ell\ell$ or $B_s \to \phi \ell \ell$ always accompanies the perturbative SM contribution and the NP's one of the Wilson coefficient:
$${\cal C}_{9\, j}^{\rm eff}=C_{9 \rm pert}^{\rm eff SM} + C_{9}^{\rm NP} + C_{9\,  j}^{\bar{c}c}$$
where j stands for the different transversity amplitudes. This charm contribution is amplitude and process dependent, while NP is universal.

In particular, it was argued in \cite{silver} the possibility that some unknown contribution coming from soft gluon interchange with loops of charm maybe causing the deviations. Of course, the statement is rather generic in these terms and an explicit and real computation would be very welcome  to prove it and not simply performing a fit as in  \cite{silver}.
In our approach we include as an additional uncertainty in the error budget the only  existing computation in literature in the framework of LCSR of soft gluon emission from four-quark operators involving $\bar{c}c$ currents \cite{KMPW}, but with one important addition. We allow  for the interference term to have any sign $C_{9\,  j}^{\bar{c}c}=s_j {\cal C}_{9 \, j}^{\rm \bar{c}c KMPW}$
with $s_i \in [-1,1]$. This is a rather drastic and conservative approach, but we consider that in absence of information on the relative sign is an appropriate approach. As a consequence of it  long-distance charm-loop is the main contributor to our error budget.

 In the following we  list five arguments to show that the possibility that huge unknown charm contributions  are behind the observed anomalies is  rather  unprobable  when confronted with more elaborated analysis or when taking a global view (see \cite{hadronic} for more details):

\begin{itemize}

\item[I)] In order to confirm or not the statement in \cite{silver} (later on amended by same authors in \cite{silver2}) we performed an analysis of $B \to K^*\mu\mu$ data in \cite{hadronic} within a frequentist approach using the same polynomial parametrization as in \cite{silver}. This parametrization  consists in adding to each amplitude a term of the form $h_\lambda=h_\lambda^{(0)}+\frac{q^2}{1 {\rm GeV}^2} h_\lambda^{(1)}+ \frac{q^4}{1 {\rm GeV}^4} h_\lambda^{(2)}$. However, notice that $h_\lambda^{(i)}$ parametrize a q$^2$-expansion of the charm-loop and obviously NP contributions leaks in all terms:
a NP contribution to $C_7$ affects  $h_\lambda^{(0)}$ but also higher orders when expanding $C_i^{\rm NP} \times FF(q^2)$, same for $C_9$ affecting $h_\lambda^{(1)}$ and higher orders. The problem reduces to answer the question if a large $q^2$ dependence arises in $C_{9 \, j}^{\rm eff}$ (i.e., a significant non-zero contribution to $h_\lambda^{(2)}$ is required besides the one that leaks from NP~\footnote{Notice that  in \cite{silver} is implied that a non-zero $h_\lambda^{(2)}$ can only be induced by a non-trivial charm contribution and not reabsorbed  in a shift of the Wilson coefficients. 
This statement misses the small contribution from the expansion mentioned above.} )   or instead data is compatible with a universal shift of $C_9$ as a global NP contribution would imply. This can be done in a systematic way by
 testing the improvement or not of the quality of the fit when going from the hypothesis of a constant contribution (NP-like contribution) to the hypothesis of a $q^2$-dependent one. We implemented different analysis:  SM scenario, NP scenario, different sets of FFs  and even including all data from $b \to s\ell\ell$. In none of the cases we analyzed in \cite{hadronic} we observed any significant improvement in the quality of the fit that pointed to the need to go beyond the $h_\lambda^{(1)}$ term in the polynomial expansion (equivalent to a constant contribution to $C_9$), implying no need of a non-trivial $q^2$-dependent contribution. Indeed after the authors of \cite{silver} agree with us in \cite{silver2} that a constant universal contribution gives indeed a good description of the data as we already found longtime ago, these same authors now speculate on the possibility that this constant is of non-perturbative origin. Besides the fact that this is based on pure speculation and not on any real computation, we see two different problems on this rather speculative (albeit  impossible to discard) possibility. First, this would require that this non-perturbative contribution must be transversity and process independent. However, explicit computations \cite{KMPW} of non-factorizable long-distance $c\bar{c}$ contributions already with one soft gluon exchange found results that are completely different in $B \to K^*\ell\ell$ (between longitudinal and the rest of amplitudes) but specially with $B \to K\mu^+\mu^-$ where they are found to be negligible. And second, even if one would find some contrived example, it would be impossible to explain $R_K$, $R_{K^*}$ and $Q_5$. Instead as shown in the next section the same New Physics that explains LFUV gives a nearly perfect prediction for $P_5^\prime$ leaving very little space  to accommodate such a contrived non perturbative unknown constant contribution to $P_5^\prime$ (see Fig.2).

\item[II)] LHCb performed an  experimental analysis~\cite{petridis}  on the mode $B \to K \mu\mu$ where they measured the relative phases of the $J/\psi$ and $\psi(2S)$ with the short-distance contribution and found small interference effects in dimuon mass region of interest far from pole masses of the resonances. The outcome was that these long-distance contributions cannot explain the observed deviations in this channel.

\item[III)] A group of experimentalists  performed also recently an  analysis~\cite{egede}  to determine the long distance contributions in $B \to K^{*0} \mu\mu$ using an empirical model of long-distance contributions 
based on the use of data on final states involving $J^{\rm PC}=1^{--}$ resonances. The outcome shown in Fig.2 shows a very good agreement with our estimates. Notice that in \cite{egede} the BSZ-FF are used while Fig.2 is recomputed in the framework of KMPW.
Besides arriving to the same conclusion that in previous case but for the mode $B \to K^*\mu\mu$ there is the added interesting information of the good agreement with our prediction for $P_5^\prime$. 

\item[IV)] Another  analysis \cite{javier} based on analyticity properties is used to fix the q$^2$ dependence up to a polynomial in $B \to K^* \ell\ell$ observables. They obtain the long-distance contributions in this mode in the region of interest for different observables  finding that the SM fits are significantly inefficient in front of a NP hypothesis to explain data in a range from 3.4 to 6$\sigma$. It is also shown that the deviation in $P_5^\prime$ is far from being explained within the SM.

\item[V)] Finally, a particularly strong argument is the fact that the LFUV observables  (Type-II)  that are not affected by long-distance contributions in the SM show deviations in very nice agreement with Type-I if NP affects mostly muons and not electrons as data seems to point to (see Fig.9 in \cite{global}).

\end{itemize}

\begin{figure}
\begin{minipage}{5cm}
\includegraphics[width=5.2cm, height=6.3cm]{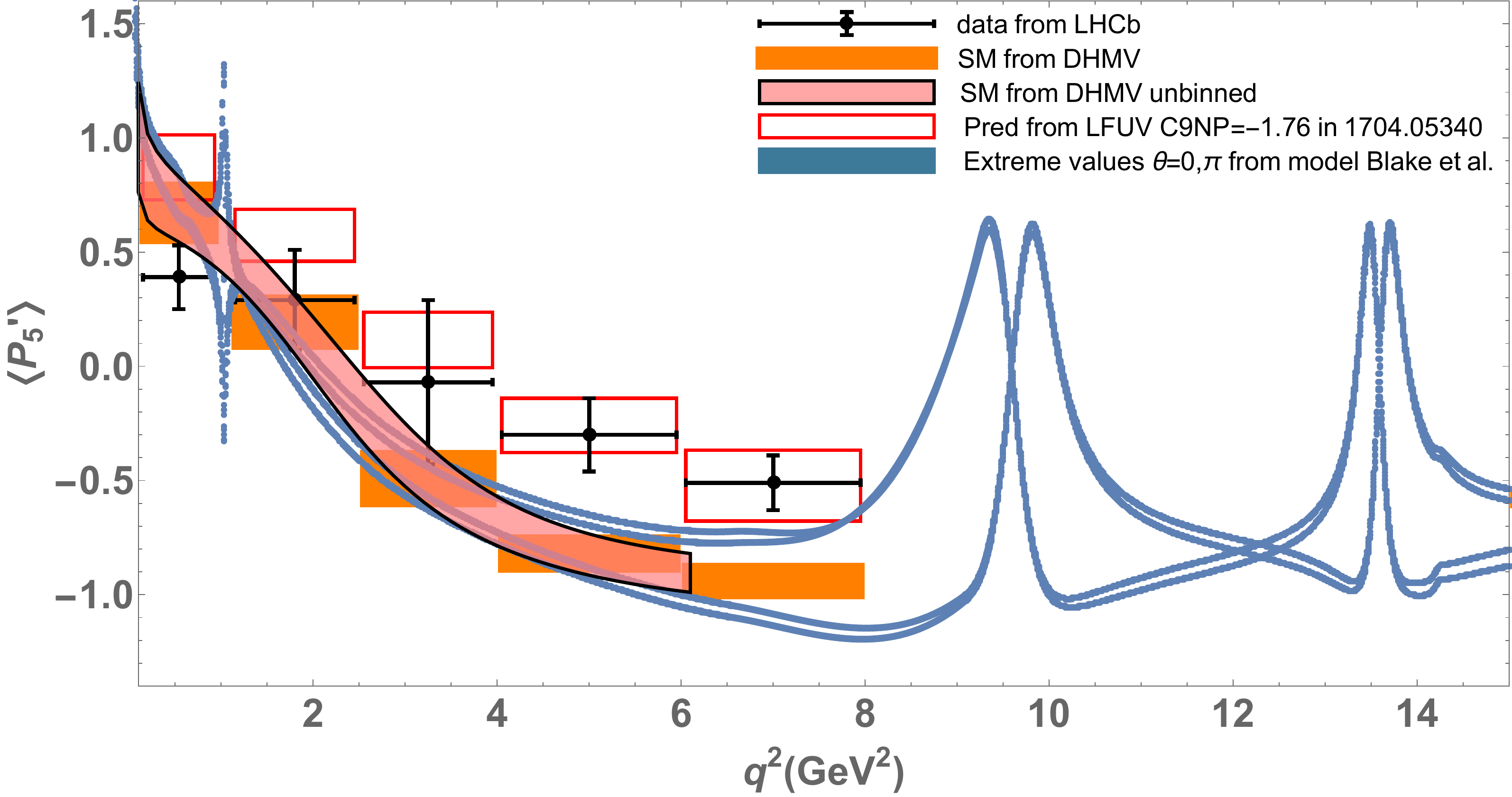}
\end{minipage}
\,\,\vspace*{-0.3cm}\begin{minipage}{5cm}\vspace*{-0.4cm}
\includegraphics[width=5.2cm, height=5.9cm]{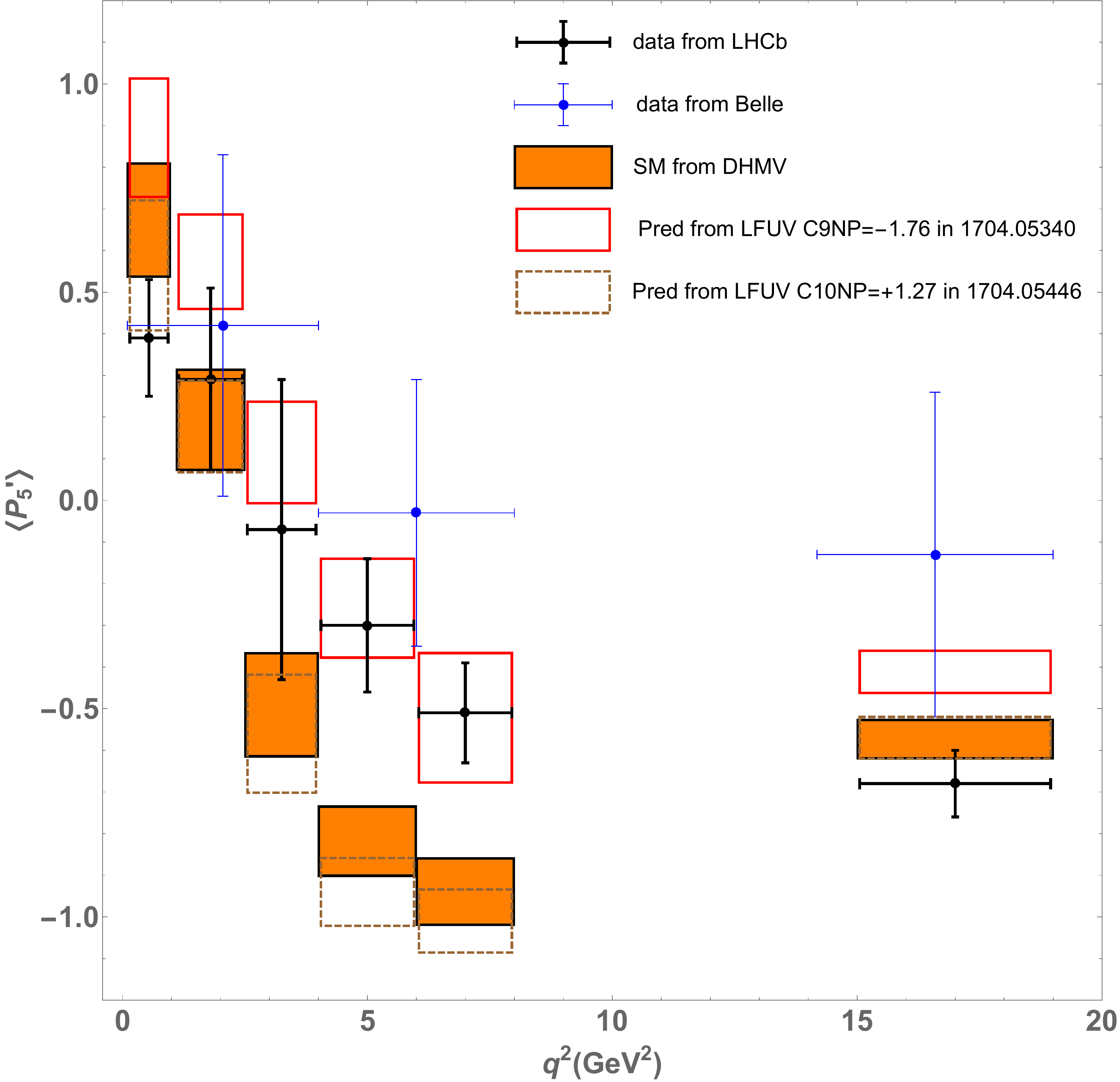}
\end{minipage}
\begin{minipage}{5cm}
\includegraphics[width=5.4cm, height=6.5cm]{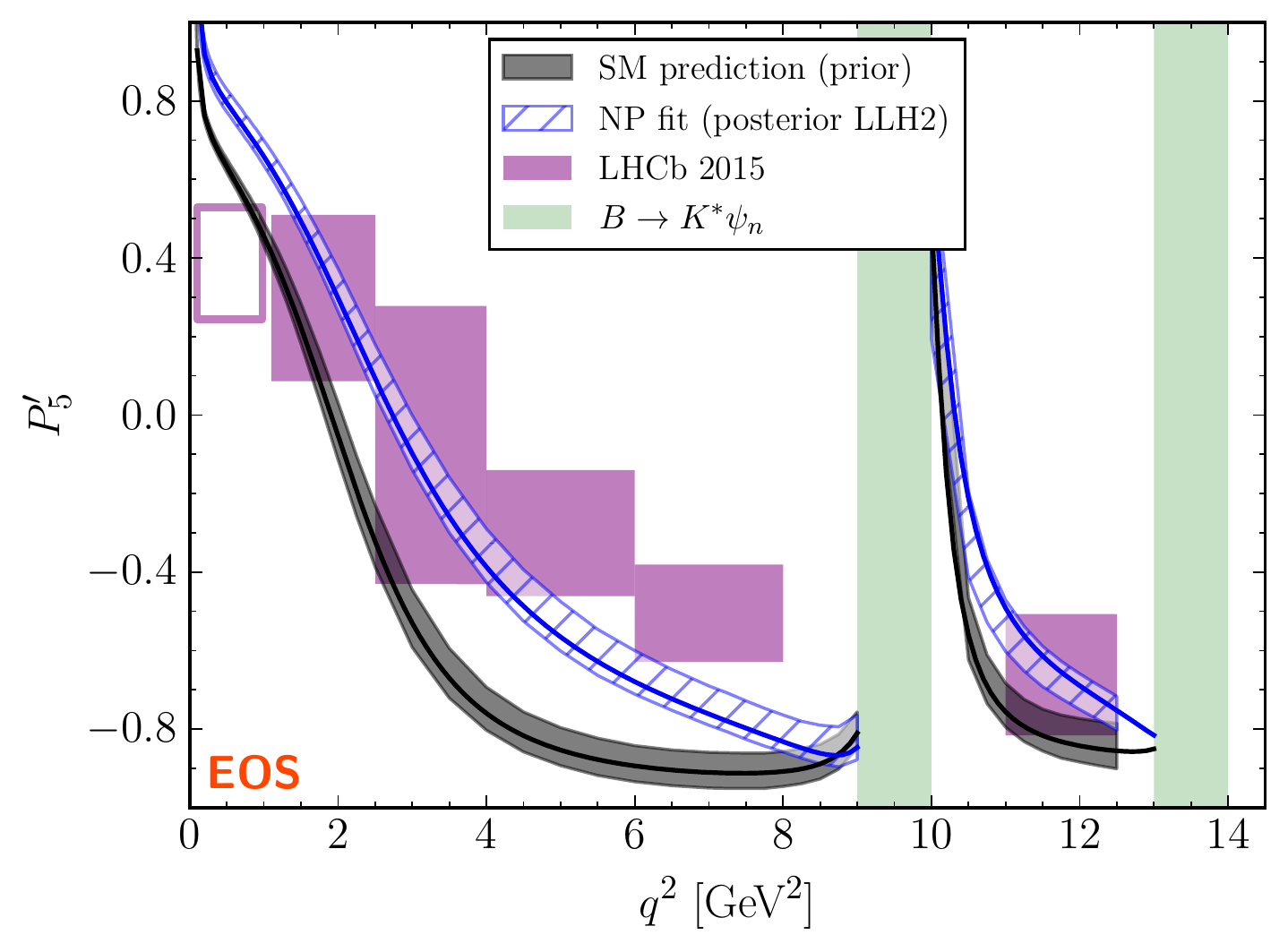}
\end{minipage}
\caption{Plots corresponding to arguments left (III) and right (IV). See details in text. Center: SM binned prediction in orange, LHCb (Belle) data in black (blue) and  in red NP prediction of $P_5^\prime$ in the scenario with NP only in $C_9$  obtained  from LFUV observables [12] and in brown for a scenario with NP only in $C_{10}$ [41].}
\end{figure}


\subsection{Hadronic uncertainties in LFUV observables}
 
 There are two types of LFUV observables, ratios of branching ratios, like $R_{K,K^{*}}$ and observables based on differences of optimized observables, like $Q_i=P_i^{\prime \mu}-P_i^{\prime e}$ or based on non-optimal ones like $T_i$ (see \cite{asessing} for definitions). All of them are equally efficient in the first step to test and quantify the presence or not of NP, when compared with their very precise predictions in the SM.  Only tiny radiative corrections can induce an uncertainty that does not cancel in the ratio (see \cite{isidori}). Still 
 the non-cancellation of terms of order $(m_\mu^2-m_e^2)/q^2$ that are practically irrelevant for $q^2 > 1$ GeV$^2$ produces a residual hadronic uncertainty in the SM coming from FFs and long distance contributions. The situation is completely different in presence of NP. Then the error associated to NP predictions for ratios of branching ratios that are totally unprotected in front of FF choices  can vary substantially  if BSZ or KMPW FF are used (see appendix in \cite{pattern}). In this sense calling {\it clean} $R_{K,K^{*}}$ in presence of NP is totally misleading. Also  estimating the error associated to soft FF  as the spread of central values of FF determinations an in \cite{jaeger2,jaeger3} can also have dramatic effects in the error associated to the predictions of $R_{K,K^{*}}$ observables in presence of NP. Instead, observables like $Q_5$ lead to robust predictions (small differences if BSZ or KMPW FF are used)
as opposite to the case of $R_{K^*}$  (see appendix of \cite{pattern} and compare $R_{K^*}$ with $Q_5$ error). This gives to $Q_5$
%
sufficient discriminating power  to discern between scenarios pointing to NP mainly in $C_9^{\rm NP}$ with minor impact in the rest of Wilson coefficients or $C_9^{\rm NP}=-C_{10}^{\rm NP}$ scenarios.
 
 Finally an important exercise is to cross-relate analyses, namely see the implications of one analysis on the observables of the other analysis. For instance,  in \cite{pattern} we asked ourselves the following question: what is the value predicted for $P_5^\prime$ from the global fit taking exclusively LFUV observables. The result is quite impressive: the LFUV observables in the scenario with NP only in $C_9$ predicts $C_{9 \, \rm LFUV}^{\rm NP}=-1.76$ and this in turn implies a {\bf prediction} for $P_5^\prime$ exactly where data measures $P_5^\prime$ to be (see Fig.2). Indeed an efficient way to discard other scenarios is to check what they would predict for $P_5^\prime$. For instance, a scenario  with NP only in $C_{10}^{\rm NP}=1.27$ (see \cite{jaeger3})
would predict for $P_5^\prime$ a value below the SM and far from data. As Fig.2 shows quite explicitly, this scenario is even worst than SM to explain data.
 
 \section{Conclusions}
 
 In summary, no significant deviations are observed from the CKM paradigm even if the inclusive/exclusive tensions in the determination of $|V_{cb}|$ and $|V_{ub}|$ still persist. On the contrary a global analysis of 175 observables of $b \to s\ell\ell$, LFUV and radiative observables shows a clear pattern of deviations with respect to the SM with a global pull above 5$\sigma$ in favour of a NP hypothesis (mainly in the Wilson coefficient $C_9^{\rm NP}$ and slightly less significantly in other scenarios like $C_9^{\rm NP}=-C_{10}^{\rm NP}$). In the very short term LHCb can have the key to discriminate the preferred scenario among the different possibilities  using the $Q_i$ observables.

\section*{\it Acknowledgements}
 
JM acknowledges financial support from the grant FPA2014-61478-EXP.

\end{document}